\documentclass[sigconf]{acmart}

\usepackage{booktabs} 
\usepackage{algorithm}
\usepackage[noend]{algpseudocode}
\usepackage{url}
\setcopyright{acmlicensed}

\begin{document}

\copyrightyear{2018} 
\acmYear{2018} 
\setcopyright{acmcopyright}
\acmConference[NOSSDAV'18]{28th ACM SIGMM Workshop on Network and Operating Systems Support for Digital Audio and Video}{June 12--15, 2018}{Amsterdam, Netherlands}
\acmBooktitle{ Proceedings of NOSSDAV'18
, June 12--15, 2018, Amsterdam, Netherlands}
\acmPrice{15.00}
\acmDOI{10.1145/3210445.3210446}
\acmISBN{978-1-4503-5772-2/18/06}

\title[Rate Control for Real-time Video Streaming]{Delay-Constrained Rate Control for Real-Time Video Streaming with Bounded Neural Network}

\author{Tianchi Huang$^{\dag*}$, Rui-Xiao Zhang$^{*}$, Chao Zhou$^{\ddagger}$, and Lifeng Sun$^{*\S}$}
\affiliation{$^{*}$ Department of Computer Science and Technology, Tsinghua University, Beijing, China}
\affiliation{$^{\ddagger}$ Beijing Kuaishou Technology Co., Ltd., China}
\affiliation{$^{\dag}$ Department of Computer Science and Technology, Guizhou University, Guizhou, China}
\affiliation{\{htc17,zhangrx17\}@mails.tsinghua.edu.cn}
\affiliation{zhouchaoyf@gmail.com, sunlf@mail.tsinghua.edu.cn}
\renewcommand{\shortauthors}{Huang et al.}
\begin{abstract}
Rate control is widely adopted during video streaming to provide both high video qualities and low latency under various network conditions. However, despite that many work have been proposed, they fail to tackle one major problem: previous methods determine a future transmission rate as a single for value which will be used in an entire time-slot, while real-world network conditions, unlike lab setup, often suffer from rapid and stochastic changes, resulting in the failures of predictions. 

In this paper, we propose a delay-constrained rate control approach based on end-to-end deep learning. The proposed model predicts future bit rate not as a single value, but as possible bit rate ranges using target delay gradient, with which the transmission delay is guaranteed.
We collect a large scale of real-world live streaming data to train our model, and as a result, it automatically learns the correlation between throughput and target delay gradient. We build a testbed to evaluate our approach. Compared with the state-of-the-art methods, our approach demonstrates a better performance in bandwidth utilization. In all considered scenarios, a range based rate control approach outperforms the one without range by 19\% to 35\% in average QoE improvement.
\end{abstract}




\maketitle
\renewcommand{\baselinestretch}{0.97} \normalsize
\section{Introduction}

Recent years have seen a rapid increase in the requirements of real-time video streaming. People publish and watch live video streaming smoothly from supported applications at any time, in any where, and under any network environments. 
Due to the complicated environment and stochastic property in various network environments, the abundance of rate control approaches have been proposed to solve the fundamental problem: how to transport video stream with higher video bitrate and lower latency. For conventional approaches, loss-based approaches\cite{handley2002tcp,752152} use packet loss to control their session. Moreover, these methods will cause delay instability.\cite{8000044}. To solve this problem, delay-based approaches\cite{rossi2010ledbat,hayes2013delay} are proposed to make the end-to-end delay converge to a target value by controlling sending rate. However, the final delay of these approaches is not constrained \cite{geng2015delay}. Model-based approaches\cite{carlucci2016analysis,kurdoglu2016real,cardwell2016bbr} aim to design a model \cite{handley2002tcp}to describe the underlying relationships by analyzing observed historical network status. However, forecasting future network throughput as a single value seems to be unreliable\cite{yoshida2013constructing,zou2015can,mao2017neural}. Meanwhile, many studies focus on describing future network throughput as a probability model\cite{yoshida2013constructing}. Nevertheless, the fixed rules assumption continues to struggle with their performance.


In our study, we aim to use a suitable range to describe the future network conditions instead of using a single value. However, we cannot quantify the value of ``suitable'' clearly, and we only know the aim of ``suitable'' means ``a range can cover the future observations but not too broad.'' Motivated by this, We try to explore the insight of network congestion via end-to-end deep learning\cite{lecun2015deep}.
Starting from this concept, we propose a delay-constrained rate control approach to constrain end-to-end delay during the session. Our approach is placed on the receiver. Upon receiving packets from the sender, the receiver feeds a message containing the next time sending rate back to the sender. The sender then changes encoder's parameters to fit it. Our method consists of two modules, the delay filter, and the rate estimator. The delay filter is a module that computes delay gradient required using previously delay gradient observations. The rate estimator is an end-to-end deep learning model which can estimate future throughput in a range by using historical observations. Due to this unique requirements, the rate estimator consists of two neural networks, the prediction network (PN) and the error estimation network (EEN). By comparing each candidate architecture, we finally propose the best architecture as the rate estimator. 

After that, we evaluate our rate control approach using a full system implementation. We collect a large corpus of data from real-world network environments as a training dataset to optimize it. 
We perform several experiments to compare the performance with previously proposed approaches and evaluate the effectiveness of range factor. The key contributions of our paper are as follows:

\begin{itemize}
\item We design and train a novel neural network architecture to predict future network status in range instead of a single value, which can be more conducive to encode video in real-time video streaming scenario.

\item We propose a delay-constrained rate control approach which outputs as a range to control the session in low latency and high bitrate under the premise of stability.
\end{itemize}


\section{Related Work}
Rate control methods have been proposed and applied about two decades. These schemes are mainly classified into three types, loss-based, delay-based and model-based\cite{wu2000transporting}.

\textbf{Loss-based: }
Loss-based approaches, for instance, TFRC\citep{handley2002tcp} and rate adaptation protocol (RAP)\cite{752152}, have been widely used in TCP congestion control, which increases bitrate till packet loss occurs. However, until packet loss occurs, latency also increases, vice versa. Thus, a bad experience will be given to the users because of the network jitter like sawtooth. Thus, using packet loss event as the control signal may cause its throughput to be unstable, especially in error-prone environments\cite{geng2015delay}.

\textbf{Delay-based: }
Some studies have also focused on delay-based approaches solving the problem of the loss-based approach. Delay-based approaches try to adjust sending rate to control the transmission delay. According to the way in which these approaches calculate delay, they can be divided into the end-to-end delay (RTT) approaches, such as TCP Vegas\cite{brakmo1995tcp}; one-way delay approaches, for instance, LEDBAT (Over UDP) and TCP-LP\citep{rossi2010ledbat,kuzmanovic2006tcp}, and delay gradient approaches\cite{carlucci2016analysis}. However, it is hard for the delay metrics to converge to the target value in some network conditions which its upper limit has always suffered a wide range of changes, such as WIFI\cite{geng2015delay,8000044}.

\textbf{Model-based: }
Model-based approaches are also proposed in recent years to control congestion such as Rebera\cite{kurdoglu2016real} and GCC (Google Congestion Control)\cite{carlucci2016analysis}. These approaches are aimed to design a model or a filter to describe the underlying relationships by previously network status observations. These model forecast future throughput as a single value. Nevertheless, due to the unstable network environments, the single value may not fully describe the future network status, especially in real-time live video streaming scenario.

\begin{figure}
    \centerline{\includegraphics[width=1.0\linewidth]{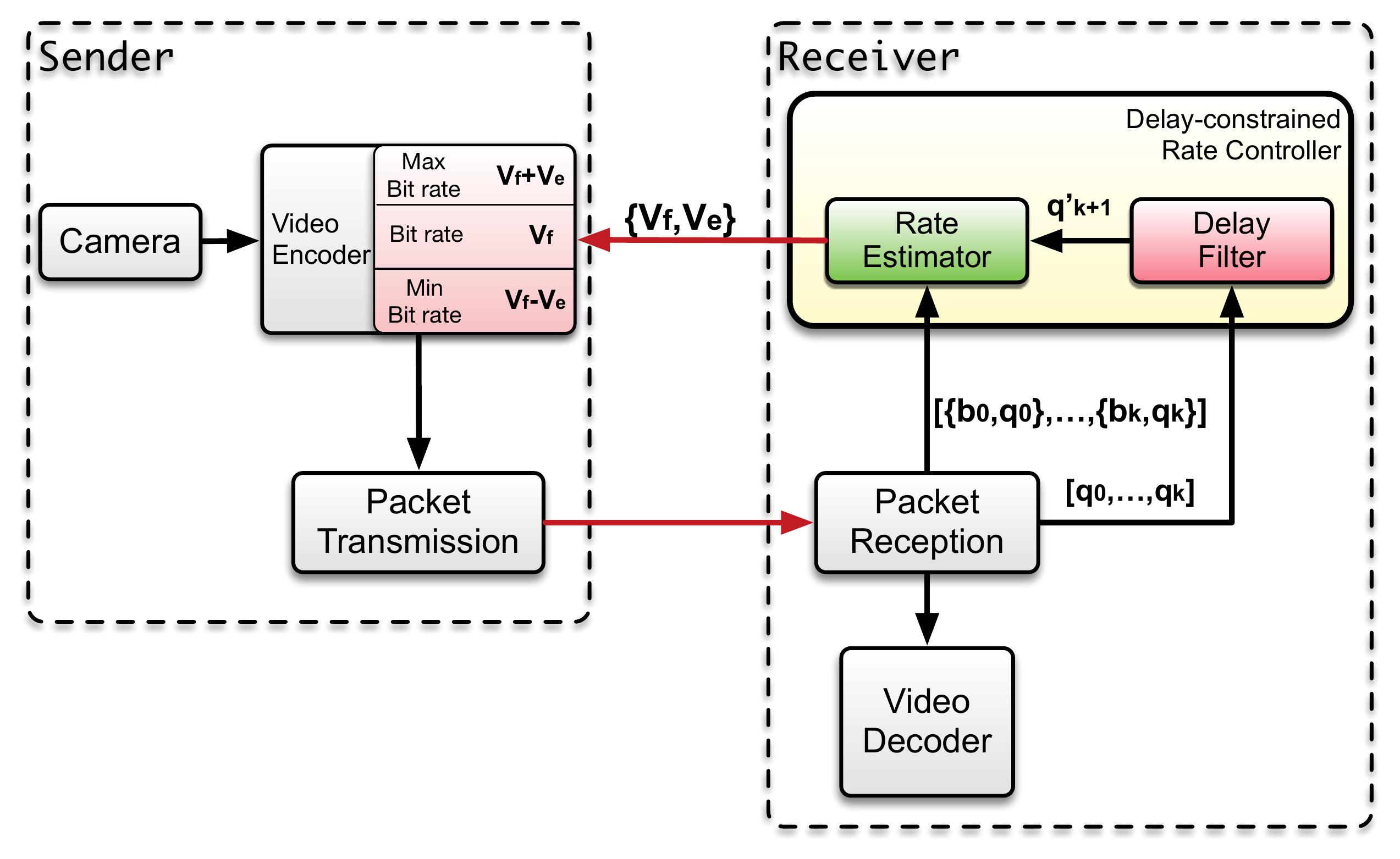}}
    \caption{Delay-constrained Rate Control System Overview}
    \label{fig:overview} 
\end{figure}

\section{System Overview}
We start with an overview of real-time live streaming scenario consisting of a sender and a receiver. Figure~\ref{fig:overview} illustrates the architecture. The sender generates the raw live streaming data using a camera and sends them to the receiver. The receiver receives the data stream and displays the view. A UDP-based protocol is established to send live video stream in MTU-sized packet\footnote{In this paper, the sender delivery of a single 1500-byte (MTU-sized) packet.} and receives feedback reports from the receiver. In general, our approach constrains the delay with the methods as follows:

\textbf{Delay filter}, placed on the receiver, which computes future delay gradient $\hat{q_{t}}$ on demand from historical delay gradient observations, collected that is fed back to the sender with the aim of constraining the delay;

\textbf{Rate estimation with bounded neural network}, placed on the receiver, which computes the sending bitrate range $[V_f - V_e , V_f + V_e]$, where $V_f$ is described as a target sending bitrate value, and $V_e$ is its error between $V_f$ and future observations;

\textbf{Rate control}, placed on the sender, which controls the encoding bitrate of the video encoder.

\section{Methods}
\subsection{Delay Filter}
Aiming to constrain the transmission delay by controlling the sending rate in the sender, we design a delay filter module that computes delay gradient on demand denoted as $\hat{q}_{k+1}$ according to delay gradient sequence $[q_0,q_1,\ldots,q_k]$ of past k time-slots. 

At the same time, when we choose the bitrate , we try to optimize the following formulation, as is shown in Eq.\ref{eq:Optimize},   
\begin{align}
 \begin{aligned}
    \label{eq:Optimize} 
     \min\left[{\overline{q}}^{2} + \alpha\sigma^{2}\right]
 \end{aligned}
 \end{align}
 
\noindent where $\overline{q}=\frac{1}{k+1}(\sum\limits_{i=1}^{k}q_{i}+\hat{q}_{k+1})$, and $\sigma^{2} = \frac{1}{k+1}(\sum\limits_{i=1}^{k}(q_{i}-\overline{q})^{2}+(\hat{q}_{k+1}-\overline{q})^{2})$. Coefficient $\alpha$ is the weight to describe its aggressiveness. T	he first part of this object tries to make the average delay gradient approximate zero to obtain the constant queuing delay and the second part is to minimize the variation of the ``change'' of the delay gradient, aiming to avoid a fierce change of queuing delay. (Eq.~\ref{eq:Controller}). We will discuss the best $\alpha$ in Section~\ref{differentalpha}.

\begin{align}
 \begin{aligned}
    \label{eq:Controller} 
       \hat{q}_{k+1} = \frac{\alpha - 1}{\alpha + k}\sum\limits_{i=1}^{k}q_{i}
 \end{aligned}
\end{align}

\begin{figure}
    \center \includegraphics[width=0.5\linewidth]{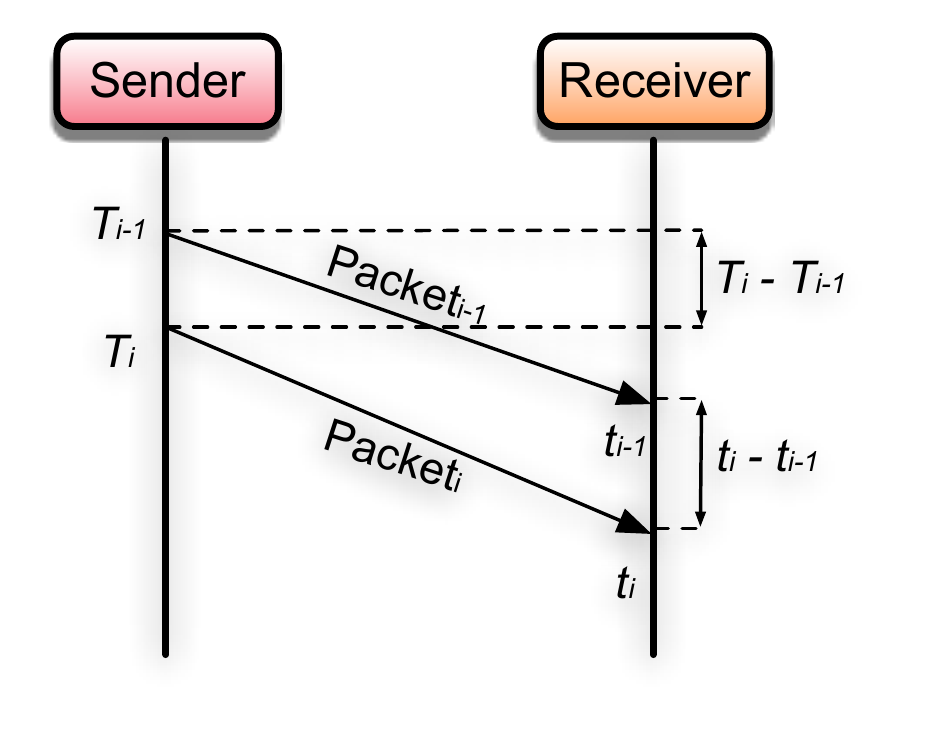}
    \vspace{-20pt}
    \caption{Delay Gradient Measurement}
    \label{fig: gradient} 
\end{figure}

\textbf{Delay Gradient Measurement.} In real-time video streaming control scenario, one of vital work for congestion control is to measure one-way delay or queuing delay of the session. However, the clocks on both sides are not synchronous which makes the delay measurement unreliable. Thus, delay gradient, defined as the difference on delay measurement on both side, has been proposed to measure the latency in unsynchronized clock environment. 

Delay gradient (Figure~\ref{fig: gradient}) $q(t_i)$ is computed as follows\cite{carlucci2016analysis} (Eq.~\ref{eq: gradient}) , which is actually used to describe the variety of the queuing delay. If the network is congested, its delay gradient will be greater than zero, vice versa. If the delay gradient equals to zero, we can only infer that network is not congested.
 \begin{align}
 \begin{aligned}
 q(t_i) = \frac{1}{N}\sum\limits_{i=1}^{N}\left[(t_i-t_{i-1})-(T_i-T_{i-1})\right]
    \label{eq: gradient} 
 \end{aligned}
 \end{align}
Where $T_i$ is the time at which the i-th packet has been sent, and $t_i$ is the time at which the i-th packet has been received, and N represents the packet count in a period. However, in real-world network environment, delay gradient is often observed along with burst noise. In our study, we filter the noise by a complementary filter.

\subsection{Rate Estimation with Bounded Neural Network}
In this section, we mainly set up the rate estimator module in several steps. We start by explaining its bounded neural network architecture including inputs, outputs, training methodology, and so forth. Compared the performance with four architecture candidates, we then confirm the best neural network architecture of rate estimator.

\subsubsection{Bounded Neural Network Architecture}

\begin{figure}
    \centerline{\includegraphics[width=0.7\linewidth]{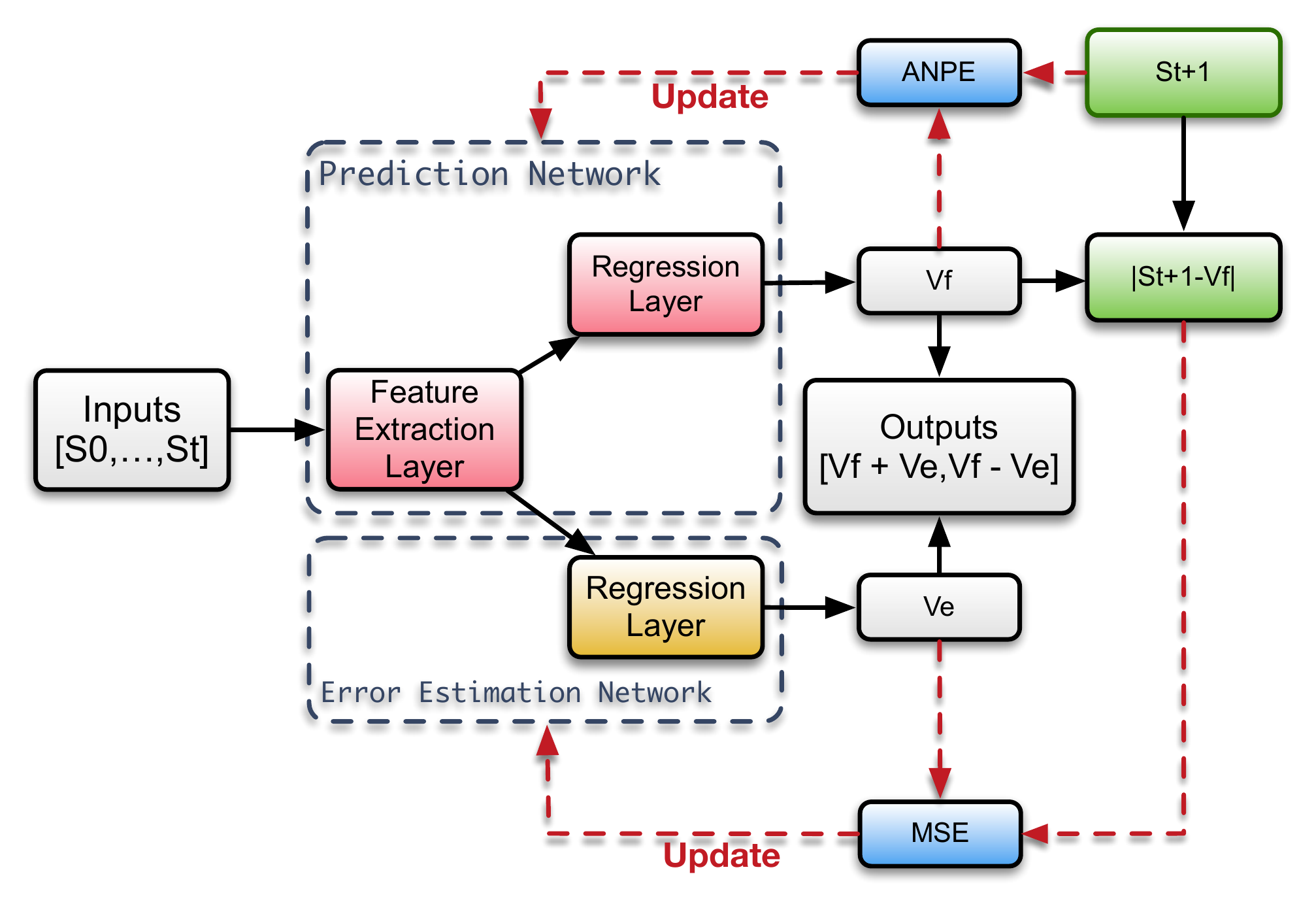}}
    \caption{Neural Network Architecture}
    \label{fig:bounded}
    \vspace{-10pt}
\end{figure}

Our motivation is to build a model which uses previously historical observation to predict the range of future observations. We model the range as $[V_f - V_e, V_f + V_e]$, where $V_f$ is described as a baseline value, and $V_e$ is its error between the baseline value and future observations. However, the range will lose its function if $V_e$ is too large. Motivated by this, We propose a novel neural network architecture, named as ''bounded neural network``, with two neural networks to maximize the accuracy using single value $V_f$ and to minimize the error value $V_e$. We describe its network structure and its training methodology. As shown in Figure~\ref{fig:bounded}, neural network uses the method which involves training two neural networks, prediction network (PN) and error estimation network (EEN). The detailed functionalities of these networks are explained below.

\textbf{Inputs}
After the duration of $k$ time-slots, rate estimator takes inputs $S_t = (b_t,q_t,e_{t+1})$ to its neural networks. Where $b_t$ is a sequence variable that represents the network throughput measurement for the past time $k$; $q_t$ is the delay gradient sequence data of the past time k, which represents the difference of each latency. Additionally, $e_{t+1}$ is the delay gradient on demand for the next time ${t+1}$, which is computed by the delay filter.

\textbf{Prediction Network}
Prediction network (PN) aims to forecast the baseline value of future observations by using historical time series observations. PN uses a neural network to make logistic regression. Furthermore, the neural network is divided into two layers, feature extraction layer, and linear regression layer. The feature extraction layer is used to extract features from inputs, and the linear regression layer is proposed to estimate the outputs with a fully connected layer. In particular, the feature extraction layer is also the inputs of EEN. For each $S$ in inputs, it is permitted to be multidimensional. 
The outputs $S_{t+1}$ of PN is a linear value that falls into $(-\infty,+\infty)$, representing the predicted value for the time ${t+1}$, which is also the baseline value of the proposed neural network.

\textbf{Error Estimation Network}
The Error Estimation Network (EEN) is used to estimate the absolute value that PN generates. The inputs of EEN is the outputs of feature extraction layer in PN. Same as PN's outputs, a linear value is proposed to describe the error in PN, which is the variable value of the proposed neural network. 

\textbf{Outputs}
Upon receiving $S_t$ , rate estimator needs to predict the value corresponding to the throughput $S_{t+1}$ for the given delay gradient $e_{t+1}$ of next time $t+1$. In our model, the value is represented by a range from minimum value to maximum value which is inflected by the output of PN and EEN, respectively.
\begin{algorithm}[b]
\caption{Neural Network Gradient Training Algorithm} 
\label{alg:gradient} 
\begin{algorithmic}[1]
\renewcommand{\algorithmicrequire}{\textbf{Input:}}
\Require 
Historical observations $[S_0,S_1, \ldots, S_t]$, \newline Prediction observations $S_{t+1}$, Learning rates $\alpha, \beta$
\Procedure{Update Gradient}{$[S_0,S_1, \ldots, S_t], S_{t+1}$}
\State $\hat{S}_{t+1} \gets PN.Predict([S_0,S_1, \ldots, S_t])$;
\State $Loss_{pred} \gets ANPE(S_{t+1},\hat{S}_{t+1})$;
\State $E_{t+1} \gets \left|\hat{S}_{t+1}-S_{t+1}\right|$;\Comment{ground truth of EEN}
\State $H_{pred} \gets PN.FeatureExtractionLayer()$;
\State $\hat{E}_{t+1} \gets EEN.Predict(H_{pred})$;
\State $Loss_{err} \gets \left|E_{t+1}-\hat{E}_{t+1}\right|^{2}$;
\State $PN \gets PN - \alpha\nabla_{pred}Loss_{pred}$
\State $EEN \gets EEN - \beta\nabla_{err}Loss_{err}$
\EndProcedure
\end{algorithmic} 
\end{algorithm}
\subsubsection{Training Methodology}
We now describe how to train and optimize the neural network architecture. The pseudo-code for training methodology is given in Algorithm \ref{alg:gradient}. PN and EEN update their gradient respectively during the training process. After receiving $S_{0 \ldots t}$, PN predicts the future observation $\hat{S}_{t+1}$. By comparing the ground truth value $S_t$, we can optimize PN. To evaluate the difference between value predicted and the ground truth, we use absolute normalized prediction error (ANPE), which is described in Eq.\ref{EQ1}. 
After calculating ANPE, PN then updates its gradients for minimizing ANPE by using back propagation method. While updating the gradient of PN, the absolute error between value predicted and true value (Eq.\ref{EQ2}), which is denoted as $E_t$, will be used as a ground truth of EEN. After that, EEN will use mean square error (MSE) to update its network gradient.
 \begin{align}
 \begin{aligned}
    ANPE(y,\hat{y})=\left|\frac{y-\hat{y}}{\hat{y}}\right|
    \label{EQ2}
 \end{aligned}
 \end{align}
 \vspace{-10pt}
 \begin{align}
 \begin{aligned}
    E_t = \left|y_t-\hat{y_t}\right|
    \label{EQ1}
 \end{aligned}
 \end{align}


\begin{figure}
    \centerline{\includegraphics[width=0.8\linewidth]{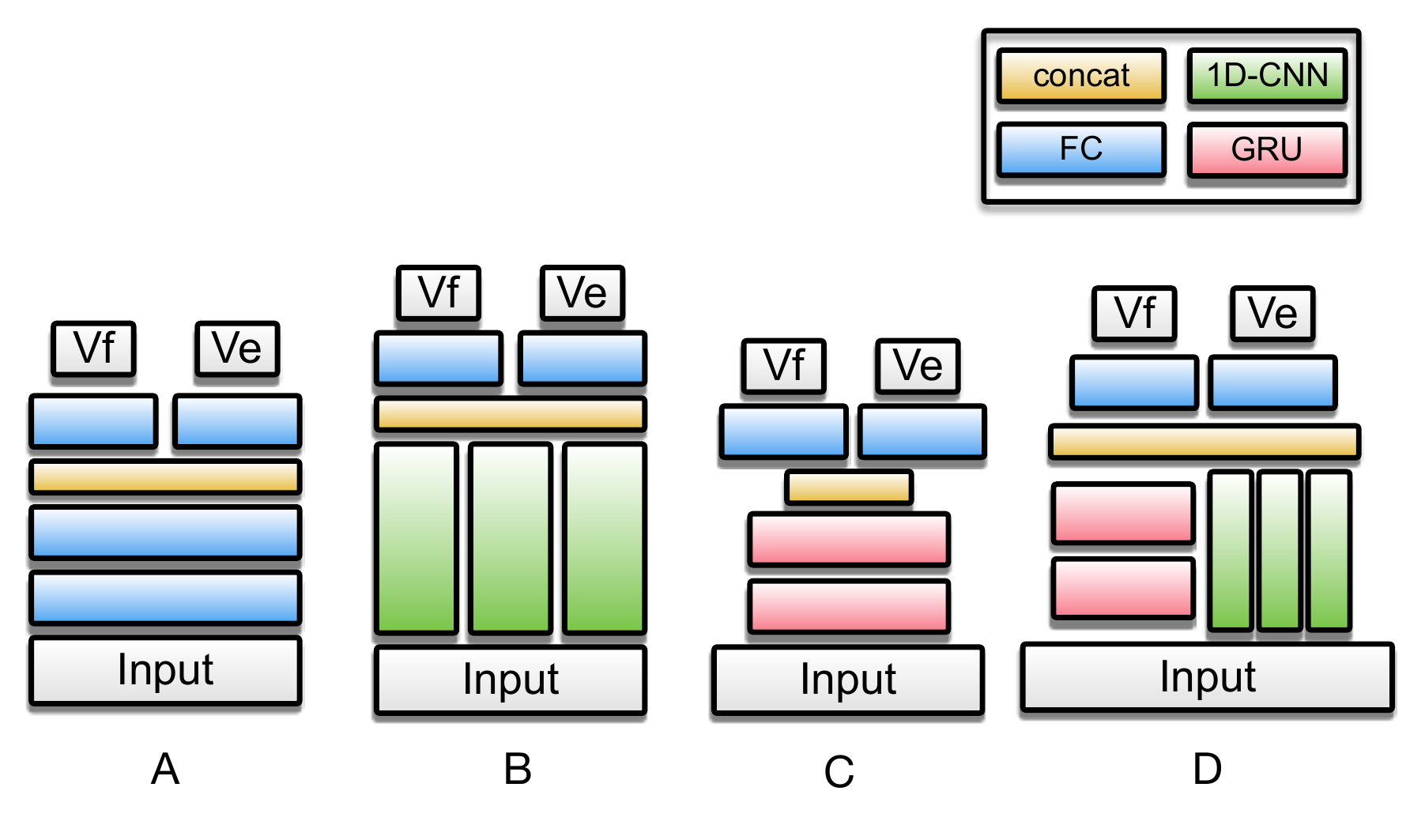}}
    \vspace{-10pt}
    \caption{Architecture Overview For Each Candidate}
    \label{fig:candidates} 
    \vspace{-10pt}
\end{figure}

\subsubsection{Best Architecture Selection} To select the best architecture, we derive four different deep neural network architectures for the rate estimator as depicted in Figure~\ref{fig:candidates}, named as Arch-A to Arch-D, respectively. Arch-A considers network variable features as a common feature, so we only use two fully-connected layers to design its architecture. Arch-B is similar except that some of network metrics are assumed as a signal input, and more internal and hidden features can be shown by convolution neural network (CNN). The next two architectures consider the temporal properties. The third design (Arch-C) uses GRU, a well-known recurrent neural network (RNN), to extract the hidden features of time series. The last design (Arch-D) is a hybrid neural network architecture which combines both Arch-A and Arch-C with a merged layer. We test and compare the prediction accuracy with four architectures for selecting the best one. We collect a small scale of network status as the dataset from real-world network environment for training and validation. In detail, we leverage ANPE (Eq.\ref{EQ1}), error estimation rate (EER) and coverage rate (CR) to analyze the estimation accuracy of each architecture. EER is defined as follows w.r.t. root mean square error (RMSE) (Eq.\ref{EER}).

\begin{align}
 \begin{aligned}
    EER = 1.0 - \sqrt{\frac{1}{n}\sum\limits_{i=1}^{n}\left| \hat{E_i} - E_i \right| ^{2}}
 \label{EER}
  \end{aligned}
\end{align}

\vspace{-5pt}
\begin{align}
 \begin{aligned}
    CR = \frac{1}{n}\sum_{i=1}^{n}{P_i}
     \label{CR},where\,P_i =
   \begin{cases}
       1 &\mbox{$\hat{y_i} \in [V_f - V_e, V_f + V_e]$}\\
    0 &\mbox{$\hat{y_i} \notin [V_f - V_e, V_f + V_e]$}
   \end{cases}
 \end{aligned}
\end{align}

Where $\hat{E_{i}}$ represents the estimated throughput, $E_{i}$ is the future throughput observed, and n is the size of test dataset. CR is employed in a ratio to estimate the coverage of throughput range during the session (Eq.~\ref{CR}). After training and testing, the final results are shown in Table \ref{T:pos2}. In short, by comparing integrated among ANPE from PN, EER from EEN and CR from range predicted of each network, we summarize their performance as follows: 
\begin{itemize}
\item Despite the outstanding performance of CR by using Arch-A, the EER metric in a small value indicates that its error estimation network has not been converged completely. Therefore, ARCH-A cannot adequately predict the error range.

\item Comparing the overall performance with Arch-B, Arch-C, and Arch-D, we find that Arch-D has better overall performance. 
\end{itemize}

Finally, we choose Arch-D, a hybrid neural network to implement the neural network in rate estimator.

\begin{table}
\begin{center}
\begin{tabular}{lp{1.5cm}p{1cm}p{1cm}p{1cm}}
\hline
\textbf{Architecture} & \textbf{ANPE} & \textbf{EER} & \textbf{CR \( \% \)} \\
\hline
Arch-A & 0.41 & 0.14 & 100.0 \\
Arch-B & 0.32 & 0.98 & 94.0 \\
Arch-C & 0.51 & 0.99 & 92.3 \\
\textbf{Arch-D} & \textbf{0.29} & \textbf{0.99} & \textbf{95.2} \\
\hline

\end{tabular}

\end{center}
\caption{Results for Bounded Neural Network Architectures}
\label{T:pos2}
\vspace{-25pt}
\end{table}

\subsubsection{Implementation} 
In conclusion, rate estimator passes $k=8$ past throughput to the feature extraction layer, which consists of a convolution network (CNN) and a recurrent network.The convolution network passes $k$ = 8 past throughput to the convolution layer with 64 filters, each of size 3 with stride 1. Past delay gradient measurement is passed to another 1D-CNN with the same shape. Results from these layers are then aggregated with other inputs in a hidden layer with 64 neurons, which can be regarded as the hyper feature of inputs. The recurrent network passes $k$ = 8 throughput and delay gradient measurement to a gated recurrent unit(GRU) layer with 64 hidden units, then the states of that layer are passed to another GRU layer with the same hidden units. A hidden layer is defined as the hidden output of the last GRU layer. These two hidden layers are finally merged into one layer, which is the inputs of EEN. Then the neural network is divided into two parts. Similarly, both parts use the final output as a linear neuron, but their duties are quite different. For PN, it updates the gradient from the network inputs to the prediction outputs. For EEN, it only updates the gradient from the extracting layer to error estimation outputs.\footnote{Details of the neural network architecture can be found at \\ \textcolor{blue}{\url{https://github.com/godka/bounded-neural-network}}.}During the training process, learning rates $\alpha,\beta$ for PN and EEN are configured to be 0.000625 and 0.001 respectively. In our experiment, we use single GPU GTX1080Ti to train our model, and our neural network converges within 100 epochs. Finally, we use TensorFlow\cite{abadi2016tensorflow} to implement this architecture, in particular, we leveraged the TFLearn deep learning library's TensorFlow API to declare PN and EEN.

\subsection{Rate Control}
We now consider how to combine the outputs of the rate estimator with the video encoder parameters. 
The video encoder requires a smooth bitrate range rather than a bitrate value, so 
we set video encoder type as constrained variable bit rate (CVBR) . In this type, encoder needs $bitrate_{max}, bitrate_{min}, bitrate$ as inputs, so after receiving the feedback message which contains the $V_f, V_e$, we set $bitrate_{min} = V_f - V_e, bitrate = V_f, bitrate_{max} = V_f + V_e$. Although the bitrate range can be controlled by changing $\alpha$ in the delay filter module. In practice, we also use a moving average threshold to protect the encoder from switching frequently in stationary network environments.

\begin{figure}
    	\includegraphics[width=0.49\linewidth]{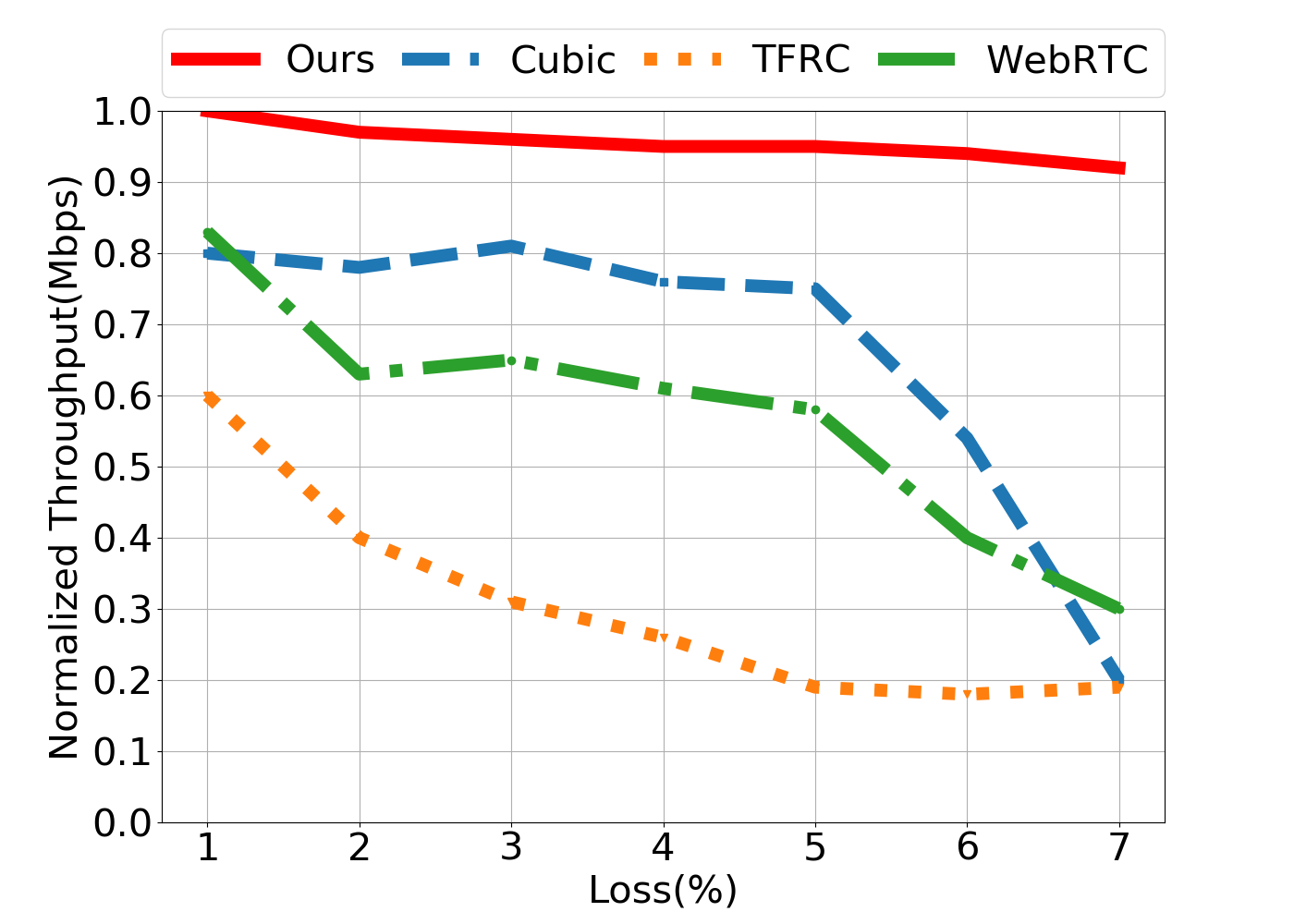}
    	\includegraphics[width=0.49\linewidth]{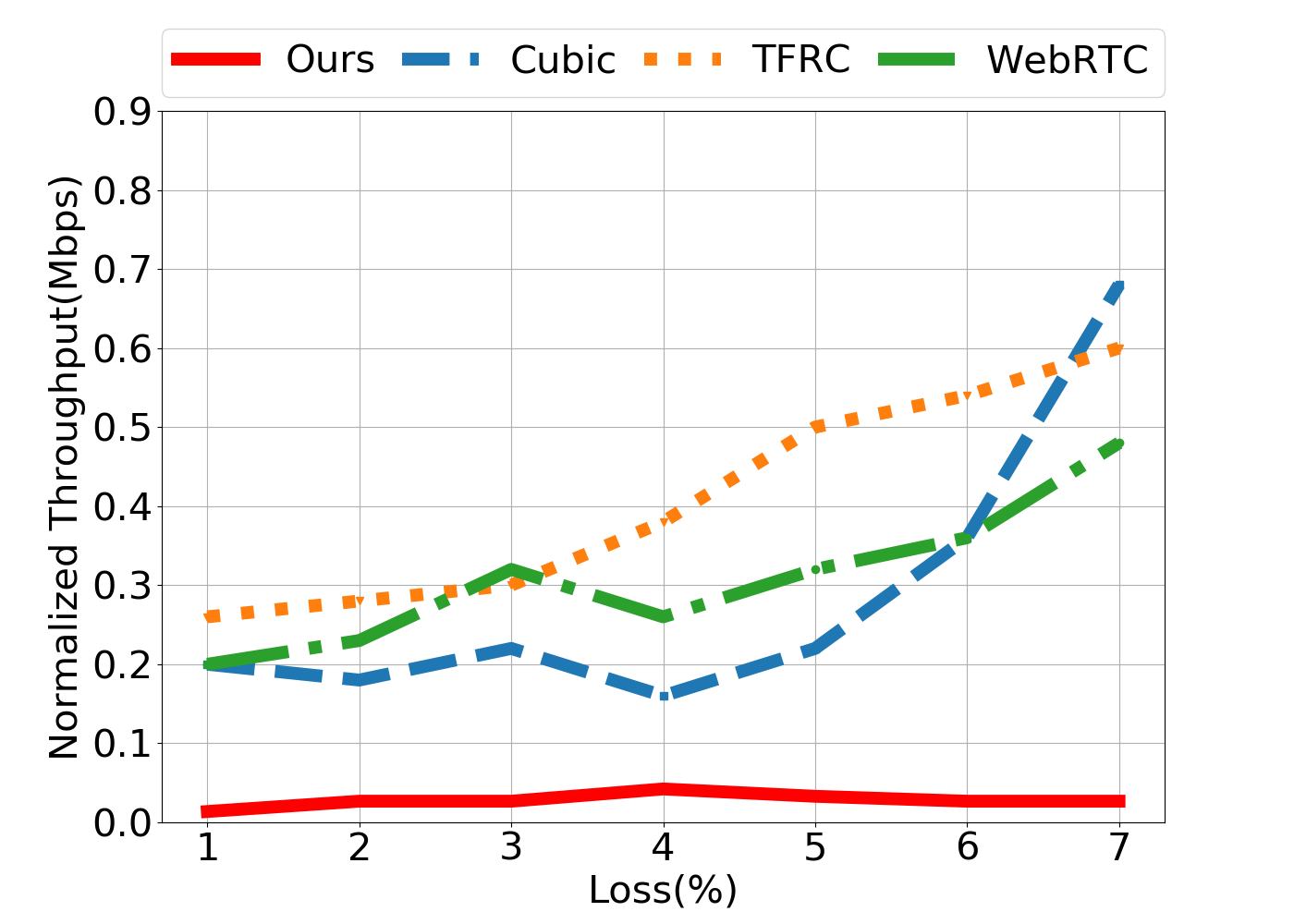}
    	\caption{Comparing our method with several proposed methods, results are collected under different loss ratio in the simulation environments.}
    	\label{T:experiments1loss2}
\end{figure}

\section{Experiments and Results}
In this section, we establish a real-time video streaming system to experimentally evaluate our rate control method. We compare our approach with current widely used methods on a wide range of network conditions. Our results answer the following questions:

\begin{enumerate}
\item What is the best coefficient $\alpha$ in our scenario?

\item By comparison of different schemes, how much impact does range have on the results?

\item Compared with previously proposed schemes, how does our approach perform on the throughput, latency, and their stability performance?
\end{enumerate}

\subsection{Dataset} To train rate estimator module, we collect a large number of network status dataset. For this reason, we use a proprietary dataset of packet-level live-cast session status from all platforms APPs of Kuaishou collected in January 2018. \footnote{Kuaishou is a leading platform in China which has over 700 million users worldwide, and millions of original videos are published on it every day. } The dataset consists of over 14 million sessions from 47,000 users covering 50 thousand unique sessions over three days in January 2018. In each session, a receiver set up a KTP (Kuaishou Transport Protocol) connection with a sender and pull video stream. Within each session, we collect network status of the stream in packet-level including send time, receive time, receive packet size, and session id.


\subsection{Simulation Environment Experiments} 
In this experiment, we aim to evaluate the average throughput and performance stability of several state-of-the-art rate control methods in some simulated network environments, and by using Network Emulator for Windows Toolkit, we can simulate different environments. After running five minutes for each approach, we collect the average throughput from the receiver. We compare their performance under different loss ratio. In this experiment, the performance is compared with TFRC, a hybrid loss-based approach\cite{handley2002tcp}, Cubic\cite{ha2008cubic}, a conventional loss-based approach and Google Congestion Control\cite{carlucci2016analysis}, a hybrid delay-based and loss-based approach that is used in WebRTC. 
Normalized throughput results have been illustrated in Figure~\ref{T:experiments1loss2}, which has demonstrated that the performance of our proposed method outperforms the widely used approaches, with improvements in average packet loss and latency of 10\% to 83\%, especially in error-prone environments, when packet loss bursts, our approach remains high throughput and stability.

\subsection{Real Testbed Experiments} 
Starting from scratch, as depicted in Figure~\ref{fig:overview} we set up a testbed where the sender uploads live stream captured by the camera to the receiver via UDP protocol. Our rate control approach is set up on the receiver. The sender has two modules: one is a video encoder which encodes video as H.265 baseline profile; the other is the packet transmission module that transfer the stream encoded to the receiver. The receiver is composed of two main parts, packet reception module, which receives video streaming from the sender and measures its network status, and rate controller, which uses hybrid neural network architecture to constrain the end-end delay. With theoretical analysis and recent experiments, we set time-slot $t$ to 1.0s, and target delay is equal to the latency that is measured from the first time-slot.\footnote{In this paper, we focus on how to get high video bitrate and bandwidth utilization with delay-constrained instead of estimating target delay.} The evaluation of rate control algorithm is split into three experiments.

\begin{table}
\begin{center}
\begin{tabular}{lp{1cm}p{1cm}p{1cm}p{1cm}p{1cm}}
\hline
\textbf{$\alpha$} & \textbf{$B_T$(\%)} & \textbf{$\sigma{(T)}$} & \textbf{$\bar{L}$(ms)} & \textbf{$\sigma{(L)}$} \\
\hline
0 & 83.3 & 0.0888 & 10.7279 & 0.6904 \\
0.6 & 84.3 & 0.0868 & 10.6047 & 0.7126 \\
\textbf{1.2} & \textbf{84.7} & 0.0840 & \textbf{10.4435} & \textbf{0.2868} \\
1.8 & 83.2 & \textbf{0.0832} & 10.7916 & 0.6046 \\
2.4 & 82.8 & 0.0867 & 10.5954 & 0.5552 \\
\hline
\end{tabular}
\end{center}
\caption{Comparing performance of the delay filter that is initialed by different coefficient $\alpha$  with throughput and latency measurement on the network emulator. Results are collected under the fixed bandwidth = 4Mbps, loss rate = 10\% and latency = 100ms respectively.}
\label{tab:widgets}
\label{T:experiments1loss}
\vspace{-25pt}
\end{table}

\subsubsection{Comparison Of Different Coefficient $\alpha$.}
\label{differentalpha}
In this part, we design an experiment to confirm the best coefficient $\alpha$ to optimize the delay filter module. We compare throughput and latency with different $\alpha$ in the same network environment. Two representative experiments are designed to evaluate our proposed approach. The result is shown in Table\ref{T:experiments1loss}, which includes bandwidth utilization and standard deviation of throughput ($B_T,\sigma{(T)}$) and latency ($\bar{ L},\sigma{(L)}$) by the delay filter. The delay filter is started with $\alpha \in \{0,0.6,1.2,1.8.2.4\}$. In our experiment, we set $\alpha$ as 1.2 to estimate next delay gradient on demand.

\begin{align}
     \begin{aligned}
    \begin{split}
        USI = &2.15 \times \log(bitrate) - \\
            &1.55 \times \log(jitter) - 0.36 \times RTT
    \end{split}
    \end{aligned}
    \label{eq:USI}
\end{align}

\subsubsection{Effectiveness of Range Factor.} 
\begin{figure}
    \begin{minipage}{1.0\linewidth}
     		\centerline{\includegraphics[width=0.83\linewidth]{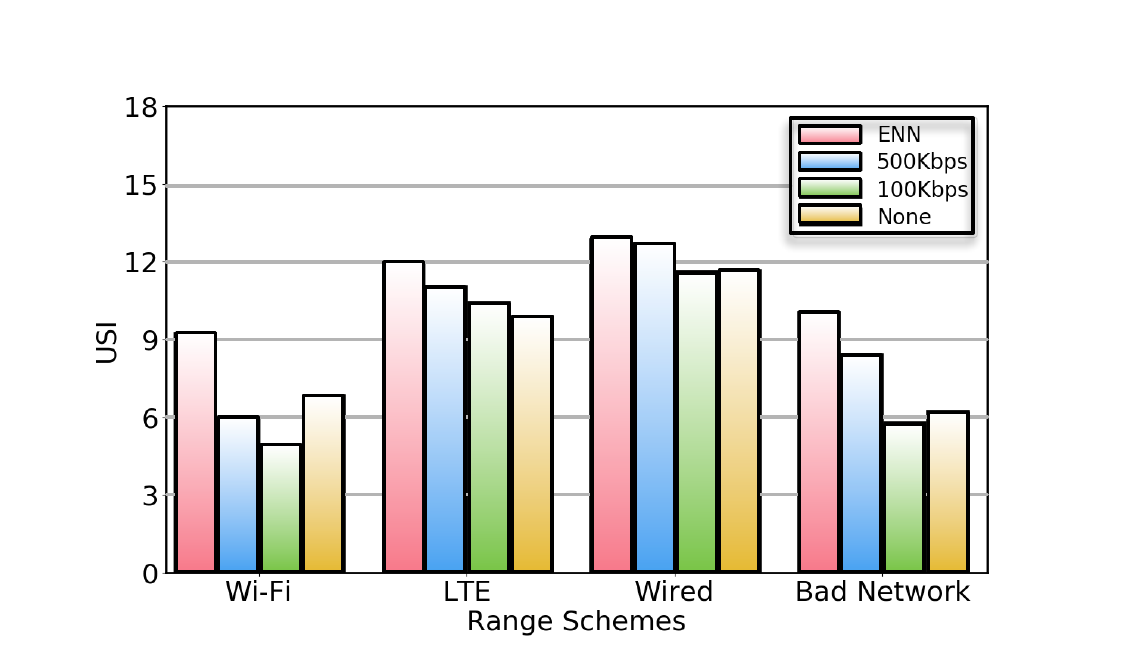}}
		\vspace{-10pt}
    		\caption{Comparing with the USI score of different ranges, results are collected under various network conditions, such as Wi-Fi, LTE, wired, and ``very bad network''.} 
    		\label{T:experrange}
    \end{minipage}
    \begin{minipage}{1.0\linewidth}
\centerline{\includegraphics[width=0.7\linewidth]{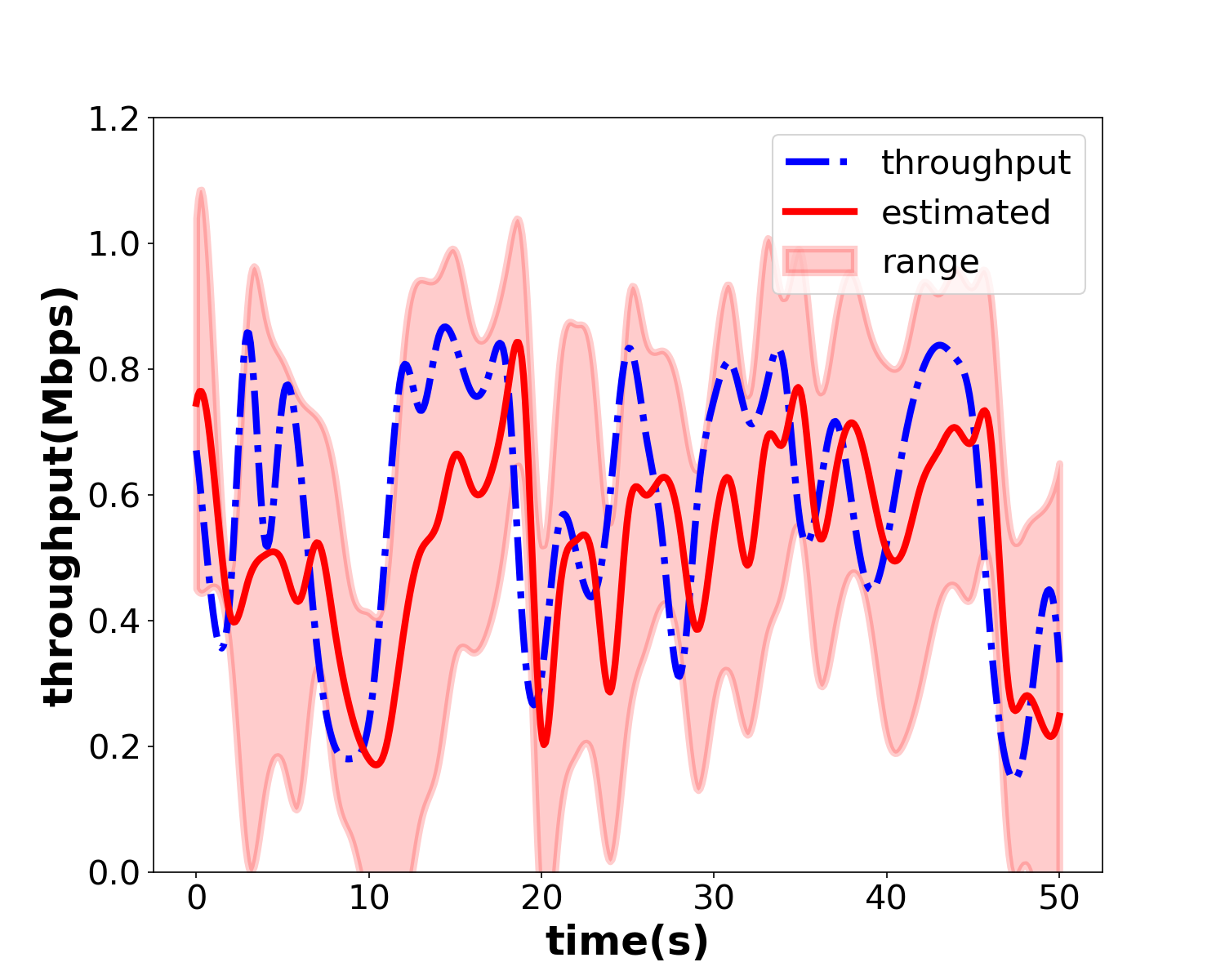}}
		\vspace{-10pt}
		\caption{Comparing the accuracy of  neural network using range with the one without using range, the result shows that estimating throughput in range can be more conducive to describe the future observations.}
		\label{fig:range} 
    \end{minipage}
\end{figure}

We evaluate the effectiveness of our range. We compare our method with the ones with fixed range and the one without range, using the Users Satisfaction Index (USI) \cite{chen2006quantifying} defined as the QoE metric (Eq.\ref{eq:USI}). Where the bitrate is the average video bitrate, jitter is the delay gradient, and RTT is the round-trip time between the sender and the receiver. We use two fixed range methods as baselines, which are 100Kbps and 500Kbps. The result is plotted at Figure~\ref{T:experrange}. From the result, we can see that dynamic range method from EEN performs better than any other range schemes. In particular, our approach using range works well in ``very bad network''\footnote{Very bad network is a network profile in Network Link Conditioner (a tool which is provided by Xcode), and describes the network status under the bandwidth = 1Mbps, Loss rate = 10\% and latency = 500ms.}, and ``Wi-Fi'' environments, which improves 35\% and 19\% in USI respectively. The output of our rate control approach is shown in Figure~\ref{fig:range}, and comparing with the one without using range, our method is more conductive. 

\section{Conclusion}
In this paper, we focus on rate control method in the real-time live streaming scenario. Previously proposed approaches are all devoted to finding a single precise prediction value which can not well adapt to the stochastic dynamics of network conditions.
To solve this open issue, we describe future observations as a range instead of a value. In this paper, we design a delay-constrained rate control approach based on end-end deep learning. The proposed method predicts a future throughput range with both previously network status and future delay required. To optimize its accuracy in different network characteristics, we train this on a large number of real-time video streaming data. In the experiment, our approach is deployed on the receiver, and by comparing with the state-of-the-art methods and the method without using range, our method all shows a better performance in bandwidth utilization. The study also concludes that predicting an optimal range performs better than predicting a value.

Additional research may focus not only on the real-time live streaming scenario but also on any time-series forecasting environments.\newline
\noindent\textbf{Acknowledgments.} We thank the anonymous NOSSDAV reviewers for their valuable feedback. This work was done in close cooperation with Kuaishou Technology Co., Ltd., and was part-funded by the National Natural Science Foundation of China under Grant No. 61472204, 61521002
, 
Beijing Key Laboratory of Networked Multimedia (Z161100005016051). 

\bibliographystyle{ACM-Reference-Format}
\bibliography{main}


\begin{thebibliography}{19}


\ifx \showCODEN    \undefined \def \showCODEN     #1{\unskip}     \fi
\ifx \showDOI      \undefined \def \showDOI       #1{#1}\fi
\ifx \showISBNx    \undefined \def \showISBNx     #1{\unskip}     \fi
\ifx \showISBNxiii \undefined \def \showISBNxiii  #1{\unskip}     \fi
\ifx \showISSN     \undefined \def \showISSN      #1{\unskip}     \fi
\ifx \showLCCN     \undefined \def \showLCCN      #1{\unskip}     \fi
\ifx \shownote     \undefined \def \shownote      #1{#1}          \fi
\ifx \showarticletitle \undefined \def \showarticletitle #1{#1}   \fi
\ifx \showURL      \undefined \def \showURL       {\relax}        \fi
\providecommand\bibfield[2]{#2}
\providecommand\bibinfo[2]{#2}
\providecommand\natexlab[1]{#1}
\providecommand\showeprint[2][]{arXiv:#2}

\bibitem[\protect\citeauthoryear{Abadi, Barham, Chen, Chen, Davis, Dean, Devin,
  Ghemawat, Irving, Isard, et~al\mbox{.}}{Abadi et~al\mbox{.}}{2016}]%
        {abadi2016tensorflow}
\bibfield{author}{\bibinfo{person}{Mart{\'\i}n Abadi}, \bibinfo{person}{Paul
  Barham}, \bibinfo{person}{Jianmin Chen}, \bibinfo{person}{Zhifeng Chen},
  \bibinfo{person}{Andy Davis}, \bibinfo{person}{Jeffrey Dean},
  \bibinfo{person}{Matthieu Devin}, \bibinfo{person}{Sanjay Ghemawat},
  \bibinfo{person}{Geoffrey Irving}, \bibinfo{person}{Michael Isard},
  {et~al\mbox{.}}} \bibinfo{year}{2016}\natexlab{}.
\newblock \showarticletitle{TensorFlow: A System for Large-Scale Machine
  Learning.}. In \bibinfo{booktitle}{\emph{OSDI}}, Vol.~\bibinfo{volume}{16}.
  \bibinfo{pages}{265--283}.
\newblock


\bibitem[\protect\citeauthoryear{Brakmo and Peterson}{Brakmo and
  Peterson}{1995}]%
        {brakmo1995tcp}
\bibfield{author}{\bibinfo{person}{Lawrence~S. Brakmo} {and}
  \bibinfo{person}{Larry~L. Peterson}.} \bibinfo{year}{1995}\natexlab{}.
\newblock \showarticletitle{TCP Vegas: End to end congestion avoidance on a
  global Internet}.
\newblock \bibinfo{journal}{\emph{IEEE Journal on selected Areas in
  communications}} \bibinfo{volume}{13}, \bibinfo{number}{8}
  (\bibinfo{year}{1995}), \bibinfo{pages}{1465--1480}.
\newblock


\bibitem[\protect\citeauthoryear{Cardwell, Cheng, Gunn, Yeganeh, and
  Jacobson}{Cardwell et~al\mbox{.}}{2016}]%
        {cardwell2016bbr}
\bibfield{author}{\bibinfo{person}{Neal Cardwell}, \bibinfo{person}{Yuchung
  Cheng}, \bibinfo{person}{C~Stephen Gunn}, \bibinfo{person}{Soheil~Hassas
  Yeganeh}, {and} \bibinfo{person}{Van Jacobson}.}
  \bibinfo{year}{2016}\natexlab{}.
\newblock \showarticletitle{BBR: Congestion-based congestion control}.
\newblock \bibinfo{journal}{\emph{Queue}} \bibinfo{volume}{14},
  \bibinfo{number}{5} (\bibinfo{year}{2016}), \bibinfo{pages}{50}.
\newblock


\bibitem[\protect\citeauthoryear{Carlucci, De~Cicco, Holmer, and
  Mascolo}{Carlucci et~al\mbox{.}}{2016}]%
        {carlucci2016analysis}
\bibfield{author}{\bibinfo{person}{Gaetano Carlucci}, \bibinfo{person}{Luca
  De~Cicco}, \bibinfo{person}{Stefan Holmer}, {and} \bibinfo{person}{Saverio
  Mascolo}.} \bibinfo{year}{2016}\natexlab{}.
\newblock \showarticletitle{Analysis and design of the google congestion
  control for web real-time communication (WebRTC)}. In
  \bibinfo{booktitle}{\emph{Proceedings of the 7th International Conference on
  Multimedia Systems}}. ACM, \bibinfo{pages}{13}.
\newblock


\bibitem[\protect\citeauthoryear{Chen, Huang, Huang, and Lei}{Chen
  et~al\mbox{.}}{2006}]%
        {chen2006quantifying}
\bibfield{author}{\bibinfo{person}{Kuan-Ta Chen}, \bibinfo{person}{Chun-Ying
  Huang}, \bibinfo{person}{Polly Huang}, {and} \bibinfo{person}{Chin-Laung
  Lei}.} \bibinfo{year}{2006}\natexlab{}.
\newblock \showarticletitle{Quantifying Skype user satisfaction}. In
  \bibinfo{booktitle}{\emph{ACM SIGCOMM Computer Communication Review}},
  Vol.~\bibinfo{volume}{36}. ACM, \bibinfo{pages}{399--410}.
\newblock


\bibitem[\protect\citeauthoryear{Geng, Zhang, Niu, Zhou, and Guo}{Geng
  et~al\mbox{.}}{2015}]%
        {geng2015delay}
\bibfield{author}{\bibinfo{person}{Yufeng Geng}, \bibinfo{person}{Xinggong
  Zhang}, \bibinfo{person}{Tong Niu}, \bibinfo{person}{Chao Zhou}, {and}
  \bibinfo{person}{Zongming Guo}.} \bibinfo{year}{2015}\natexlab{}.
\newblock \showarticletitle{Delay-constrained rate control for real-time video
  streaming over wireless networks}. In \bibinfo{booktitle}{\emph{Visual
  Communications and Image Processing (VCIP), 2015}}. IEEE,
  \bibinfo{pages}{1--4}.
\newblock


\bibitem[\protect\citeauthoryear{Ha, Rhee, and Xu}{Ha et~al\mbox{.}}{2008}]%
        {ha2008cubic}
\bibfield{author}{\bibinfo{person}{Sangtae Ha}, \bibinfo{person}{Injong Rhee},
  {and} \bibinfo{person}{Lisong Xu}.} \bibinfo{year}{2008}\natexlab{}.
\newblock \showarticletitle{CUBIC: a new TCP-friendly high-speed TCP variant}.
\newblock \bibinfo{journal}{\emph{ACM SIGOPS operating systems review}}
  \bibinfo{volume}{42}, \bibinfo{number}{5} (\bibinfo{year}{2008}),
  \bibinfo{pages}{64--74}.
\newblock


\bibitem[\protect\citeauthoryear{Handley, Floyd, Padhye, and Widmer}{Handley
  et~al\mbox{.}}{2002}]%
        {handley2002tcp}
\bibfield{author}{\bibinfo{person}{Mark Handley}, \bibinfo{person}{Sally
  Floyd}, \bibinfo{person}{Jitendra Padhye}, {and} \bibinfo{person}{J{\"o}rg
  Widmer}.} \bibinfo{year}{2002}\natexlab{}.
\newblock \bibinfo{booktitle}{\emph{TCP friendly rate control (TFRC): Protocol
  specification}}.
\newblock \bibinfo{type}{{T}echnical {R}eport}.
\newblock


\bibitem[\protect\citeauthoryear{Hayes and Ros}{Hayes and Ros}{[n. d.]}]%
        {hayes2013delay}
\bibfield{author}{\bibinfo{person}{David~Andrew Hayes} {and}
  \bibinfo{person}{David Ros}.} \bibinfo{year}{[n. d.]}\natexlab{}.
\newblock \showarticletitle{Delay-based congestion control for low latency}.
\newblock


\bibitem[\protect\citeauthoryear{Kurdoglu, Liu, Wang, Shi, Gu, and
  Lyu}{Kurdoglu et~al\mbox{.}}{2016}]%
        {kurdoglu2016real}
\bibfield{author}{\bibinfo{person}{Eymen Kurdoglu}, \bibinfo{person}{Yong Liu},
  \bibinfo{person}{Yao Wang}, \bibinfo{person}{Yongfang Shi},
  \bibinfo{person}{ChenChen Gu}, {and} \bibinfo{person}{Jing Lyu}.}
  \bibinfo{year}{2016}\natexlab{}.
\newblock \showarticletitle{Real-time bandwidth prediction and rate adaptation
  for video calls over cellular networks}. In
  \bibinfo{booktitle}{\emph{Proceedings of the 7th International Conference on
  Multimedia Systems}}. ACM, \bibinfo{pages}{12}.
\newblock


\bibitem[\protect\citeauthoryear{Kuzmanovic and Knightly}{Kuzmanovic and
  Knightly}{2006}]%
        {kuzmanovic2006tcp}
\bibfield{author}{\bibinfo{person}{Aleksandar Kuzmanovic} {and}
  \bibinfo{person}{Edward~W Knightly}.} \bibinfo{year}{2006}\natexlab{}.
\newblock \showarticletitle{TCP-LP: low-priority service via end-point
  congestion control}.
\newblock \bibinfo{journal}{\emph{IEEE/ACM Transactions on Networking (TON)}}
  \bibinfo{volume}{14}, \bibinfo{number}{4} (\bibinfo{year}{2006}),
  \bibinfo{pages}{739--752}.
\newblock


\bibitem[\protect\citeauthoryear{LeCun, Bengio, and Hinton}{LeCun
  et~al\mbox{.}}{2015}]%
        {lecun2015deep}
\bibfield{author}{\bibinfo{person}{Yann LeCun}, \bibinfo{person}{Yoshua
  Bengio}, {and} \bibinfo{person}{Geoffrey Hinton}.}
  \bibinfo{year}{2015}\natexlab{}.
\newblock \showarticletitle{Deep learning}.
\newblock \bibinfo{journal}{\emph{nature}} \bibinfo{volume}{521},
  \bibinfo{number}{7553} (\bibinfo{year}{2015}), \bibinfo{pages}{436}.
\newblock


\bibitem[\protect\citeauthoryear{Mao, Netravali, and Alizadeh}{Mao
  et~al\mbox{.}}{2017}]%
        {mao2017neural}
\bibfield{author}{\bibinfo{person}{Hongzi Mao}, \bibinfo{person}{Ravi
  Netravali}, {and} \bibinfo{person}{Mohammad Alizadeh}.}
  \bibinfo{year}{2017}\natexlab{}.
\newblock \showarticletitle{Neural adaptive video streaming with pensieve}. In
  \bibinfo{booktitle}{\emph{Proceedings of the Conference of the ACM Special
  Interest Group on Data Communication}}. ACM, \bibinfo{pages}{197--210}.
\newblock


\bibitem[\protect\citeauthoryear{Nihei, Yoshida, Kai, Kanetomo, and
  Satoda}{Nihei et~al\mbox{.}}{2017}]%
        {8000044}
\bibfield{author}{\bibinfo{person}{K. Nihei}, \bibinfo{person}{H. Yoshida},
  \bibinfo{person}{N. Kai}, \bibinfo{person}{D. Kanetomo}, {and}
  \bibinfo{person}{K. Satoda}.} \bibinfo{year}{2017}\natexlab{}.
\newblock \showarticletitle{QoE maximizing bitrate control for live video
  streaming on a mobile uplink}. In \bibinfo{booktitle}{\emph{2017 14th
  International Conference on Telecommunications (ConTEL)}}.
  \bibinfo{pages}{91--98}.
\newblock
\urldef\tempurl%
\url{https://doi.org/10.23919/ConTEL.2017.8000044}
\showDOI{\tempurl}


\bibitem[\protect\citeauthoryear{Rejaie, Handley, and Estrin}{Rejaie
  et~al\mbox{.}}{1999}]%
        {752152}
\bibfield{author}{\bibinfo{person}{R. Rejaie}, \bibinfo{person}{M. Handley},
  {and} \bibinfo{person}{D. Estrin}.} \bibinfo{year}{1999}\natexlab{}.
\newblock \showarticletitle{RAP: An end-to-end rate-based congestion control
  mechanism for realtime streams in the Internet}. In
  \bibinfo{booktitle}{\emph{INFOCOM '99. Eighteenth Annual Joint Conference of
  the IEEE Computer and Communications Societies. Proceedings. IEEE}},
  Vol.~\bibinfo{volume}{3}. \bibinfo{pages}{1337--1345 vol.3}.
\newblock
\showISSN{0743-166X}
\urldef\tempurl%
\url{https://doi.org/10.1109/INFCOM.1999.752152}
\showDOI{\tempurl}


\bibitem[\protect\citeauthoryear{Rossi, Testa, Valenti, and Muscariello}{Rossi
  et~al\mbox{.}}{2010}]%
        {rossi2010ledbat}
\bibfield{author}{\bibinfo{person}{Dario Rossi}, \bibinfo{person}{Claudio
  Testa}, \bibinfo{person}{Silvio Valenti}, {and} \bibinfo{person}{Luca
  Muscariello}.} \bibinfo{year}{2010}\natexlab{}.
\newblock \showarticletitle{LEDBAT: The New BitTorrent Congestion Control
  Protocol.}. In \bibinfo{booktitle}{\emph{ICCCN}}. \bibinfo{pages}{1--6}.
\newblock


\bibitem[\protect\citeauthoryear{Wu, Hou, and Zhang}{Wu et~al\mbox{.}}{2000}]%
        {wu2000transporting}
\bibfield{author}{\bibinfo{person}{Dapeng Wu}, \bibinfo{person}{Yiwei~Thoms
  Hou}, {and} \bibinfo{person}{Ya-Qin Zhang}.} \bibinfo{year}{2000}\natexlab{}.
\newblock \showarticletitle{Transporting real-time video over the Internet:
  Challenges and approaches}.
\newblock \bibinfo{journal}{\emph{Proc. IEEE}} \bibinfo{volume}{88},
  \bibinfo{number}{12} (\bibinfo{year}{2000}), \bibinfo{pages}{1855--1877}.
\newblock


\bibitem[\protect\citeauthoryear{Yoshida, Satoda, and Murase}{Yoshida
  et~al\mbox{.}}{2013}]%
        {yoshida2013constructing}
\bibfield{author}{\bibinfo{person}{Hiroshi Yoshida}, \bibinfo{person}{Kozo
  Satoda}, {and} \bibinfo{person}{Tutomu Murase}.}
  \bibinfo{year}{2013}\natexlab{}.
\newblock \showarticletitle{Constructing stochastic model of TCP throughput on
  basis of stationarity analysis}. In \bibinfo{booktitle}{\emph{Global
  Communications Conference (GLOBECOM), 2013 IEEE}}. IEEE,
  \bibinfo{pages}{1544--1550}.
\newblock


\bibitem[\protect\citeauthoryear{Zou, Erman, Gopalakrishnan, Halepovic, Jana,
  Jin, Rexford, and Sinha}{Zou et~al\mbox{.}}{2015}]%
        {zou2015can}
\bibfield{author}{\bibinfo{person}{Xuan~Kelvin Zou}, \bibinfo{person}{Jeffrey
  Erman}, \bibinfo{person}{Vijay Gopalakrishnan}, \bibinfo{person}{Emir
  Halepovic}, \bibinfo{person}{Rittwik Jana}, \bibinfo{person}{Xin Jin},
  \bibinfo{person}{Jennifer Rexford}, {and} \bibinfo{person}{Rakesh~K Sinha}.}
  \bibinfo{year}{2015}\natexlab{}.
\newblock \showarticletitle{Can accurate predictions improve video streaming in
  cellular networks?}. In \bibinfo{booktitle}{\emph{Proceedings of the 16th
  International Workshop on Mobile Computing Systems and Applications}}. ACM,
  \bibinfo{pages}{57--62}.
\newblock


\end{thebibliography}
\end{document}